\documentclass[proof]{pasj01}
\Received{$\langle$reception date$\rangle$}
\Accepted{$\langle$acception date$\rangle$}
\Published{$\langle$publication date$\rangle$}
\begin{document}
\title{Molecular clouds at the eastern edge of radio nebula W50}
\author{Haruka \textsc{Sakemi}$^{1,*}$, Mami \textsc{Machida}$^{2}$, Hiroaki \textsc{Yamamoto}$^{3}$, and Kengo \textsc{Tachihara}$^{3}$}%
\altaffiltext{1}{Graduate School of Science and Engineering, Kagoshima University, 1-21-35 Korimoto, Kagoshima, Kagoshima 890-0065, Japan}
\altaffiltext{2}{National Astronomical Observatory of Japan, 2-21-1 Osawa, Mitaka, Tokyo 181-8588, Japan}
\altaffiltext{3}{Graduate School of Science, Nagoya University, Furo-cho, Chikusa-ku, Nagoya, Aichi 464-8602, Japan}
\email{sakemi.haruka@astrophysics.jp}
\KeyWords{{ISM: individual (W50)}$_1$ --- {ISM: jets and outflows}$_2$ --- {ISM: molecules}$_3$ --- {stars: individual (SS 433)}$_4$}
\maketitle
\begin{abstract}
Microquasar SS 433 located at the geometric center of radio nebula W50 is a suitable source for investigating the physical process of how galactic jets affect the surrounding interstellar medium (ISM). Previous studies have searched for evidence of the interaction between the SS 433 jet and ISM, such as neutral hydrogen gas and molecular clouds; however, it is still unclear which ISM interacts with the jet. We looked for new molecular clouds that possibly interact at the terminal of the SS 433 eastern jet using the Nobeyama 45-m telescope and the Atacama Submillimeter Telescope Experiment (ASTE). We identified two molecular clouds, comprising many small clumps, in the velocity range of 30.1--36.5 km s$^{-1}$ for the first time. These clouds have complex velocity structures, and one of them has a density gradient toward SS 433. Although it is difficult to conclude the relation between the molecular clouds and the SS 433/W50 system, there is a possibility that the eastern structure of W50 constructed by the SS 433 jet swept up tiny molecular clumps drifting in the surroundings and formed the molecular clouds that we identified in this study.
\end{abstract}
\section{Introduction}
\label{sec:intro}
Galactic X-ray binary (GXB) jets inject matter and energy into their surroundings and thus change the physical and chemical states of the interstellar medium (ISM). The study of the interaction between the jets and surrounding ISM is thus essential to understanding the evolution of our galaxy. One of the essential contributions of the jets is the formation of molecular clouds. \citet{asahina2014} and \citet{asahina2017} carried out a magnetohydrodynamical simulation to investigate the effect of jets on surrounding neutral hydrogen (HI). In their simulation, jets compressed gas and triggered a cooling instability, and ultimately, the jets formed molecular clouds after $\sim$10$^{6}$ yr. Other studies suggested that GXB jets  drive interstellar turbulence, produce high-energy cosmic rays, and sometimes stimulate star formation \citep{cooper2020,heinz2008,mirabel2015}. Recent observational studies confirmed the interaction between galactic microquasar jets and the surrounding ISM. \citet{tetarenko2018} studied molecular clouds located on the jet axis of GRS 1915+105 using observation datasets of the Atacama Large Millimeter/sub-millimeter Array (ALMA) and found that the jet drives a shock at the impact site of the molecular cloud. Additionally, \citet{tetarenko2020} observed density and temperature tracers from molecular clouds around GRS 1758-258 and 1E 1740.7-2942 and investigated the jet feedback to the surroundings. Nevertheless, there still have been few observations of the interaction between jets and the ISM. It is thus important to increase the number of sample observations to clarify what happens in the region of interaction between jets and the ISM.

\begin{figure*}[t]
\begin{center}
	\includegraphics[width=14cm, bb=0 0 450 251]{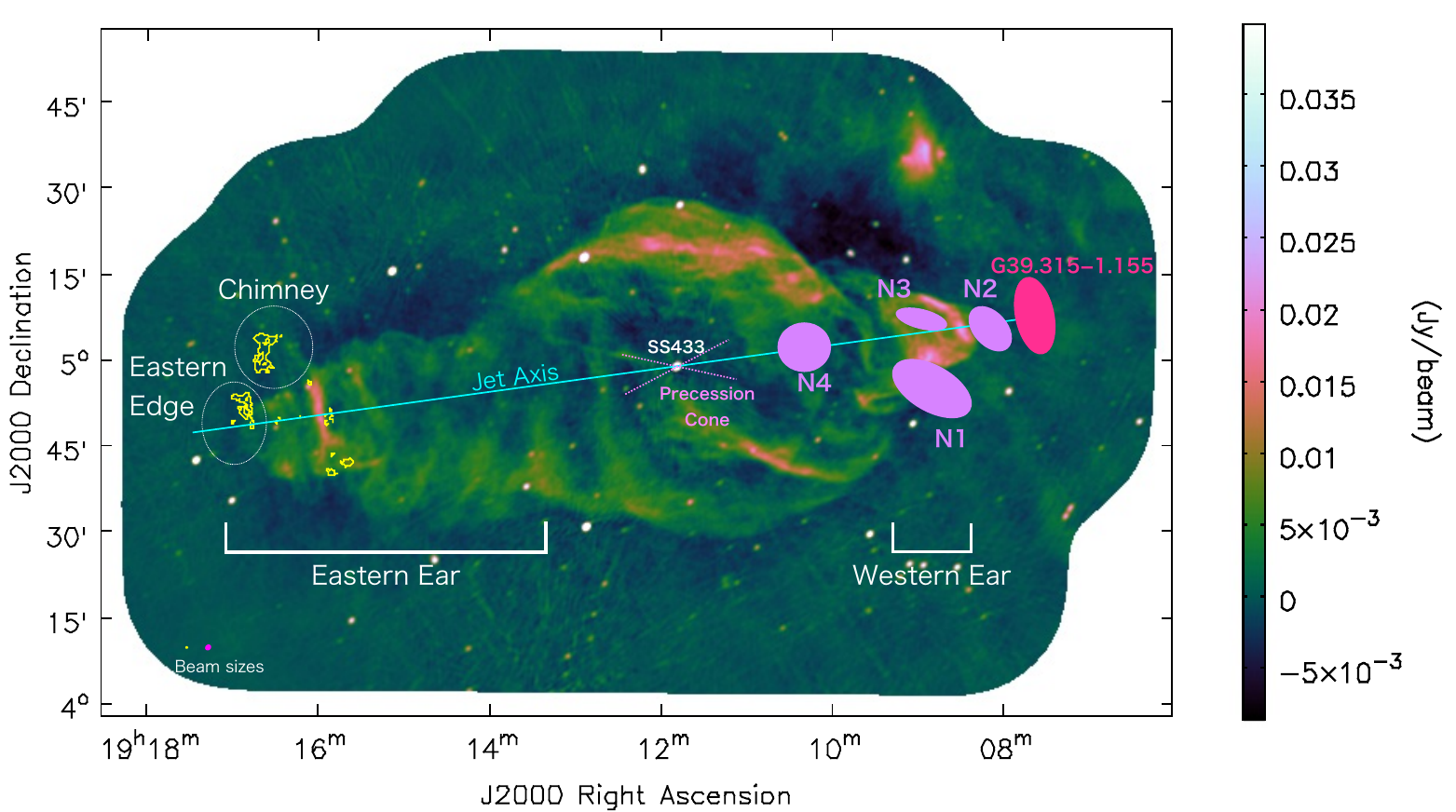}
    \caption{Radio continuum image of W50 taken with the Karl G. Jansky Very Large Array (JVLA) at 1.602 GHz \citep{sakemi2021}. Yellow contours show the positions of the newly identified molecular clouds in the velocity range 30.1--42.3 km s$^{-1}$. The contour levels are the rms $\times$ [4, 8]. The yellow circle and magenta ellipse in the bottom-left corner show the beam sizes of the CO emission and continuum, respectively. Purple and hot pink ellipses show the positions of the molecular clouds identified by previous studies in the velocity range of 49--56 km s$^{-1}$ and 73--77 km s$^{-1}$, respectively. Note that the clouds S1--S6 and G40.331--4.302 are outside the field of view.}
    \label{fig:W50}
\end{center}
\end{figure*}

The system of the microquasar SS 433 and surrounding radio nebula W50 is a suitable target for investigating the effects of GXB jets on their surroundings. W50 is a large radio nebula located at ($\alpha_{J2000}$, $\delta_{J2000}$) = (19$^{h}$ 12$^{m}$ 19.92$^{s}$, +4$^{d}$ 55$^{m}$ 01.2$^{s}$) (figure \ref{fig:W50}). It extends over an area of 2$^{\circ}$ $\times$ 1$^{\circ}$ in the sky \citep{geldzahler1980}. The microquasar SS 433 is located at the geometric center of W50. SS 433 ejects precessing jets in the east--west direction with a jet velocity of 0.26$c$, an opening angle of 20$^{\circ}$ and a precession period of 162 days \citep{abell1979,hjellming1981,margon1984,margon1989,davydov2008}. The elongated structures of W50, called ``ears'', are believed to be the result of an interaction between the jets and the nebula \citep{downes1981a,downes1981b,downes1986,dubner1998}. The western side of W50 is near the Galactic plane, and the length of the western ear along the jet axis is shorter than that of the eastern ear (19 versus 82 pc assuming a distance of 5.5 kpc). The distance of the SS 433/W50 system is still under discussion but believed to be in the range of 3.0 to 5.5 kpc.

Many studies of the ISM around SS 433/W50 have tried to search the evidence of the jet-ISM interaction and determine the kinematic distance of the system. The first observation of HI gas was reported by \citet{dubner1998} with the Green Bank Telescope. They identified an HI cavity at the position of W50 for a velocity of $v_{\rm LSR}$ = 42 km s$^{-1}$. Additionally, a large HI clump was detected at the southeast of the eastern ear, and it has been suggested that this HI clump collided with the eastern ear and changed the morphology. A distance to the SS 433/W50 system of 3.0 kpc was determined from these results. \citet{sakemi2021} compared the spatial distributions of W50 and the surrounding HI using GALFA-HI survey datasets taken with the 305-m Arecibo Radio Telescope \citep{peek2011} and confirmed that the HI cavity identified by \citet{dubner1998} has a clear spatial correlation with W50 in the velocity range from 33 to 55 km s$^{-1}$. Meanwhile, \citet{lockman2007} observed the absorption line of HI in the direction of SS 433 and concluded that the HI cloud at a velocity of 75 km s$^{-1}$ is related to the SS 433/W50 system. The velocity is clearly different from that suggested by \citet{dubner1998}, and the kinematic distance is 5.5 kpc. The distance is consistent to the value derived by comparing a deep-integrated radio image of the SS 433 jet with the kinematic jet model based on the speed and the precession period of the jet \citep{blundell2004}.

Not only the HI gas but also molecular clouds have been identified around the SS 433/W50 system. We summarize the previous studies in table \ref{table:prev-MC} and show the positions of the clouds in figure \ref{fig:W50}. \citet{yamamoto2008} suggested that the molecular clouds in the velocity range from 42 to 56 km s$^{-1}$ are related to the SS 433 jet, which is distributed around the jet axis (S1--S6, N1--N4). The eastern and western clouds of SS 433 are in the velocity range of 42 to 45 km s$^{-1}$ and 49 to 56 km s$^{-1}$, respectively. These velocities correspond to kinematic distances of 3.0 and 3.5 kpc for the flat rotation curve \citep{brand1993}. The eastern clouds (S1--S6) are distributed in a region far from the eastern edge of W50; i.e., at approximately 0.4$^{\circ}$--1.6$^{\circ}$ (outside area of figure \ref{fig:W50}). The western-side clouds (N1--N4) are closer to SS 433, and \citet{liu2020} observed N1, N2, and N3 with higher spatial resolution and concluded that a part of them is located within W50. Recently, \citet{yamamoto2022} reported the observation of N4. The molecular cloud has an asymmetric spectrum for CO emission and an extreme temperature gradient. They thus suggested that the cloud is interacting with the SS 433 jet. Meanwhile, \citet{su2018} found molecular clouds in another velocity range of 73--84 km s$^{-1}$, which are located near the western ear (G39.315--1.155) and a straight extension of the eastern ear about 0.9$^{\circ}$ (G40.331--4.302, outside area of figure \ref{fig:W50}). They suggested that the molecular clouds located at a distance of 4.9 kpc and interacted with W50 approximately 10$^{5}$ years ago. The molecular clouds identified in these previous studies have good spatial correlation with the SS 433/W50 system, and some have spectral features (e.g., spectral broadening and wings) suggesting interaction with the jet. It remains unclear at which velocities and distances the association is real, and there is no consensus in explaining this situation.

In the present paper, we focus on the eastern edge of W50. Since there is a terminal region of the SS 433 jet, it is a suitable target to investigate the jet-ISM interaction. We made observations with the Nobeyama 45-m telescope for $^{12}$CO($J$=1--0) and $^{13}$CO($J$=1--0) and with the Atacama Submillimeter Telescope Experiment (ASTE) for $^{12}$CO($J$=3--2) to identify molecular clouds for the first time. We investigate the features of the molecular clouds that we identified and provide the fundamental physical parameters based on the observed CO line emissions, such as density and temperature. The remainder of the paper is structured as follows. Section 2 describes the observations and the data reduction. Section 3 presents the observational results. Section 4 presents a discussion of the results. Section 5 provides our conclusions.

\begin{table}
 \tbl{The molecular clouds around the SS 433/W50 system identified by previous studies\footnotemark[$*$]}{%
\begin{tabular}{lccc}
 \hline
  \multicolumn{1}{c}{Name} & $V_{\rm LSR}$ (km s$^{-1}$) & Distance (kpc) & References\\
 \hline
  S1 & 42.9 & 3.0 & 1 \\
  S2 & 45.4 & 3.0 & 1 \\
  S3 & 44.1 & 3.0 & 1 \\
  S4 & 43.2 & 3.0 & 1 \\
  S5 & 44.8 & 3.0 & 1 \\
  S6 & 42.1 & 3.0 & 1 \\
  N1 & 55.8 & 3.5 & 1,2,3 \\
  N2 & 53.7 & 3.5 & 1,2,3 \\
  N3 & 53.0 & 3.5 & 1,2,3 \\
  N4 & 49.4 & 3.5 & 1,3 \\
  G39.315--1.155 & 73 & 4.9 & 4 \\
  G40.331--4.302 & 74 & 4.9 & 4 \\
 \hline
\end{tabular}}\label{table:prev-MC}
\begin{tabnote}
 \footnotemark[$*$] Column 1: Name of each molecular cloud. Column 2: Peak velocity. Column 3: Kinematic distance referred to in each paper. Note that the values of N1 to N4 refer to reference 1. Column 4: References (1)\citet{yamamoto2008}, (2)\citet{liu2020}, (3)\citet{yamamoto2022}, (4)\citet{su2018}.
\end{tabnote}
\end{table}

\section{Observations}
\label{sec:observation}
\subsection{$^{12}$CO($J$=1--0) and $^{13}$CO($J$=1--0)} 
\label{subsec:nobeyama}
The $^{12}$CO($J$=1--0) and $^{13}$CO($J$=1--0) data were obtained with the 45-m telescope of the Nobeyama Radio Observatory. We scanned an area of 30.1 $\times$ 32.6 arcmin$^{2}$ in on-the-fly mapping mode \citep{sawada2008}. We scanned in the right ascension and declination directions to suppress the scanning effect. The four-beam receiver FOREST \citep{minamidani2016} and the autocorrelation spectrometer SAM45 \citep{kuno2011} were used. Typical noise temperatures of the system including the atmosphere were between 300 and 400 K at 115 GHz. The bandwidth and resolution were 62.5 MHz and 30.52 kHz, respectively. The pointing accuracy was checked every 2 hours by observing R Aquilae ($\alpha_{J2000}$, $\delta_{J2000}$) = (19$^{h}$ 06$^{m}$ 22.25$^{s}$, +8$^{d}$ 13$^{m}$ 48.0$^{s}$) with the frontend H40 and confirmed to be better than 3 arcsec. All observations were conducted in equatorial coordinates. We used a chopper wheel to obtain the antenna temperature $T^{*}_{a}$ \citep{kutner1981}. We observed W51 as a standard source to fix gain variations between the eight output signals on December 6 to 20, 2019, February 7 to March 6, 2020, and December 1 to 4, 2020. The spectral intensity was calibrated and converted to the $T_{\rm MB}$ scale by applying a main beam efficiency $\eta_{\rm MB}$ corresponding to each polarization, frequency, and observing season provided by the observatory. The final angular resolutions were 20.3 and 20.5 arcsec at 115 and 110 GHz, respectively. The spatial and velocity grids had sizes of 8.5 arcsec and 0.2 km s$^{-1}$, respectively. The velocity coverage was from -45 to 113 km s$^{-1}$. The root-mean-square (rms) noise level in $T_{\rm MB}$ was 0.60 and 0.26 K at 115 and 110 GHz, respectively.
\subsection{$^{12}$CO($J$=3--2)}
\label{subsec:aste}
The $^{12}$CO($J$=3--2) data were obtained with the ASTE on August 9 to 11, 2019  \citep{ezawa2004}. We adopted the on-the-fly mode, and the mapping area had dimensions of 11 $\times$ 11 arcmin$^{2}$ centered at ($\alpha_{J2000}$, $\delta_{J2000}$) = (19$^{h}$ 16$^{m}$ 47.5243$^{s}$, +4$^{d}$ 51$^{m}$ 31.709$^{s}$). The frontend was a 2SB SIS mixer receiver called DASH 345. The typical system temperature was 263 K in the single side band. We used WHSF as the backend in F-FX mode. The band width and resolution were 64 MHz and 31.25 kHz, respectively. The pointing accuracy was checked every 3--4 hours by observing R Aquilae. We convolved the intensity scale into $T_{\rm MB}$ by assuming the W44 peak to be $T_{\rm MB}$ = 35.5 K \citep{wang1994}. The final angular resolution, grid size, and velocity resolution were 28.2 arcsec, 11 arcsec, and 0.2 km s$^{-1}$, respectively. The rms noise level was 0.09 K at 345 GHz.
\section{Results}
\label{sec:results}
We report the molecular cloud distribution observed with the Nobeyama 45-m telescope and ASTE at the eastern edge region of W50. Additionally, we explain the velocity features and the line-intensity ratios of the clouds.
\subsection{Spatial distributions}
\begin{figure}[t]
\begin{center}
	\includegraphics[width=8cm, bb=0 0 419 333]{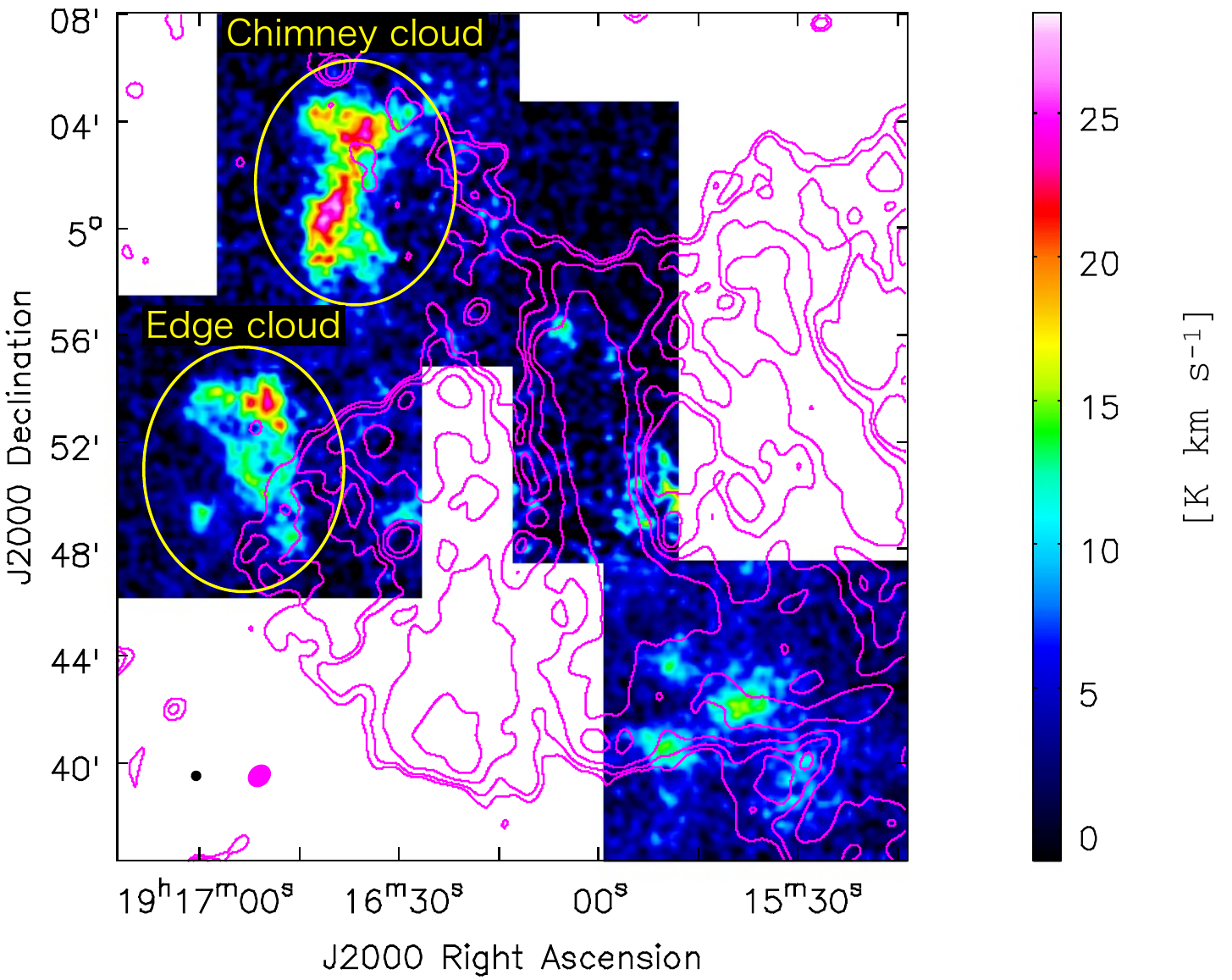}
    \caption{Velocity-integrated intensity maps of $^{12}$CO($J$=1--0). The velocity range is 30.1--42.3 km s$^{-1}$. Magenta contours show the radio continuum observed with the JVLA at 1.602 GHz. The contour levels are the rms $\times$ [2, 4, 8, 16]. The beam sizes of CO emission and continuum observations are shown in the bottom-left corner of each panel by a black circle and magenta ellipse, respectively.}
    \label{fig: integ_wide-vel}
\end{center}
\end{figure}

We show the distribution of the $^{12}$CO($J$=1--0) line emission in the velocity range of 30.1--42.3 km s$^{-1}$ observed with the Nobeyama 45-m telescope in figure \ref{fig:W50} with yellow contours. They distribute around a bright filamentary structure of W50 at $\alpha_{J2000}$ = 19$^{h}$ 16$^{m}$ 00.0$^{s}$. In this region, we found a pointed structure to the north in the radio continuum image, called the chimney (see figure \ref{fig:W50}, \cite{dubner1998,farnes2017,broderick2018}). We identified a molecular cloud near the chimney. Additionally, there was another cloud of similar size at the eastern edge of the ear. Figure \ref{fig: integ_wide-vel} is a magnified view of the target region. The color is the integrated intensity of $^{12}$CO($J$=1--0) in the velocity range 30.1--42.3 km s$^{-1}$. These velocities correspond well to those of HI gas, suggesting a relation with W50 \citep{dubner1998,sakemi2021}. We refer to the newly identified prominent clouds as the chimney cloud and edge cloud. Besides these, faint and clumpy clouds are distributed around the filamentary structure.
\begin{figure*}[t]
\begin{center}
	\includegraphics[width=16cm, bb=0 0 1308 463]{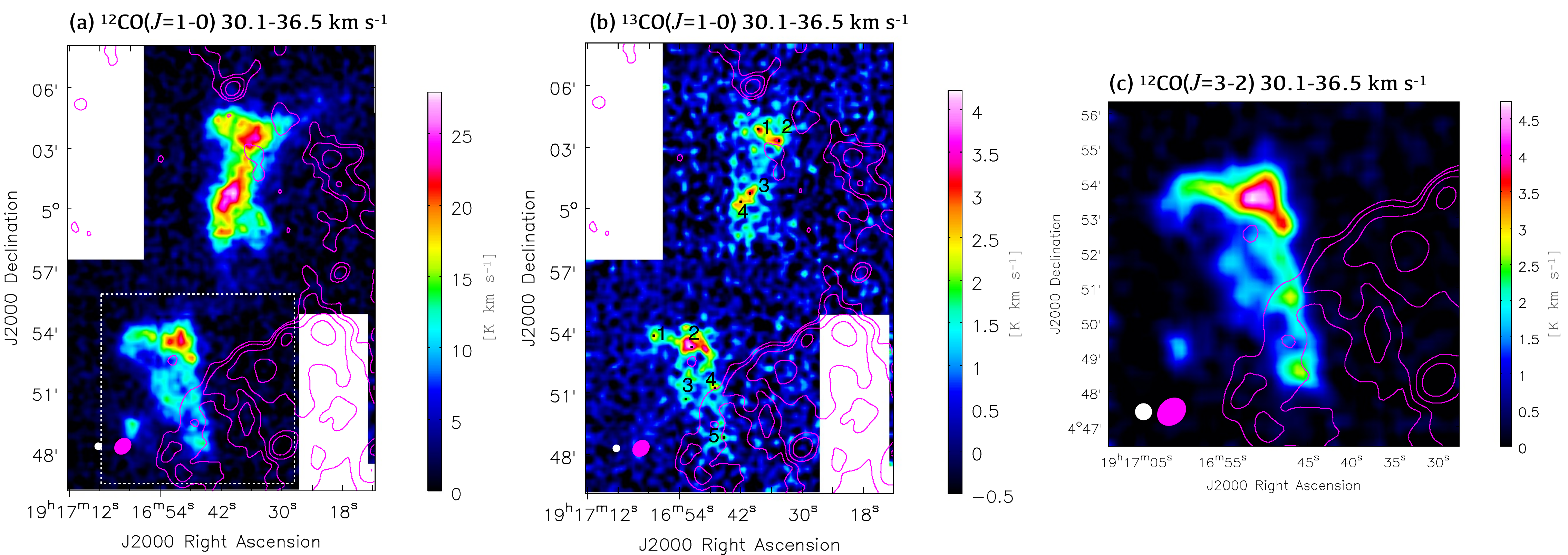}
    \caption{Velocity-integrated intensity maps of $^{12}$CO($J$=1--0) (left), $^{13}$CO($J$=1--0) (middle), and $^{12}$CO($J$=3--2) emissions (right). The velocity range is 30.1--36.5 km s$^{-1}$. Magenta contours show the radio continuum observed with the JVLA at 1.602 GHz. The beam sizes of CO emission and continuum observations are shown in the bottom-left corner of each panel by a white circle and magenta ellipse, respectively. The numbers in the middle panel show the position where the physical parameters are derived in section \ref{sec:discussion}. See table \ref{tab:column_densities} and \ref{tab:radex_results} and figure \ref{fig:LVG_edge}.}
    \label{fig:integ_30.1-36.5}
\end{center}
\end{figure*}
\begin{figure}[t]
\begin{center}
\includegraphics[width=8cm, bb=0 0 381 411]{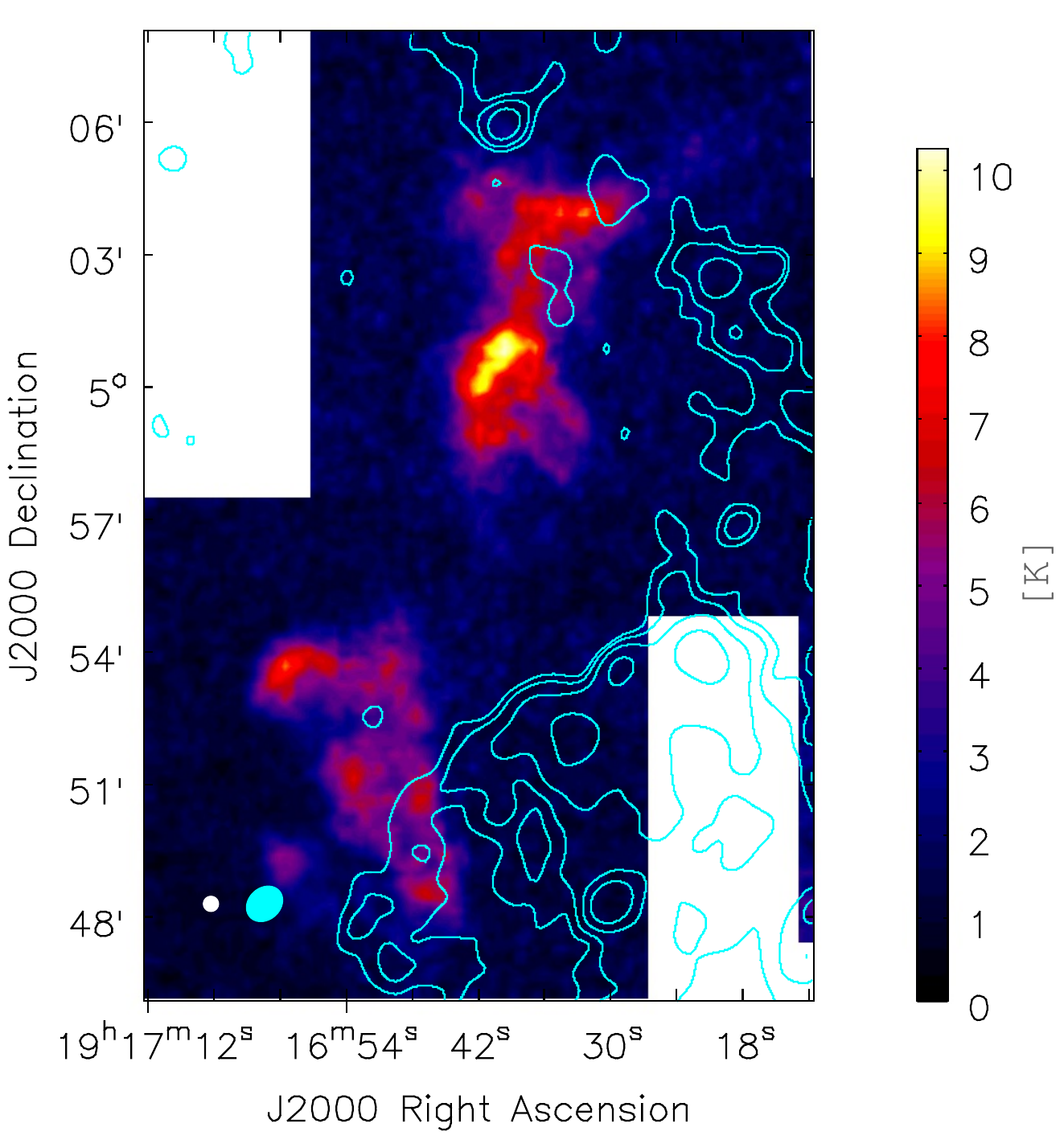}
\caption{Peak $T_{\rm MB}$ distribution in $^{12}$CO($J$=1--0). The velocity range is 30.1--36.5 km s$^{-1}$. Cyan contours show the radio continuum observed with the JVLA at 1.602 GHz. The beam sizes of CO emission and continuum observations are shown in the bottom-left corner of each panel by a white circle and cyan ellipse, respectively.}
\label{fig:peak-temperature}
\end{center}
\end{figure}

Hereafter, we focus on the chimney and edge clouds. Panel (a) of figure \ref{fig:integ_30.1-36.5} is a velocity-integrated intensity map of $^{12}$CO($J$=1--0) in the velocity range 30.1--36.5 km s$^{-1}$. The chimney cloud has two peaks in the southern and northern parts. The edge cloud is patchy, and it is brightest in the northwest region. Panel (b) of figure \ref{fig:integ_30.1-36.5} is a velocity-integrated intensity map of $^{13}$CO($J$=1--0) in the same velocity range. The intensity distribution traces the high-density regions of observed molecular clouds. Although the distribution is similar to that of $^{12}$CO($J$=1--0), the integrated intensity map of $^{13}$CO($J$=1--0) shows structures that are more clumpy. We thus expect that both the chimney and edge clouds comprise many separated clumps. Panel (c) of figure \ref{fig:integ_30.1-36.5} is a velocity-integrated intensity map of $^{12}$CO($J$=3--2). Note that we only observed the edge cloud. We detected a difference from the trend in $^{12}$CO($J$=1--0) in that the intensity was higher westward rather than eastward in the edge cloud. This cloud thus has different physical properties on eastern and western sides. 

We derived the peak $T_{\rm MB}$ distribution in $^{12}$CO($J$=1--0) (figure \ref {fig:peak-temperature}). The northern part of the chimney cloud has a flat temperature distribution. Similar to the distribution of the velocity-integrated intensity, there is a temperature peak at the southern part. In the case of the edge cloud, the temperature peak is located eastward of the northern part. This position does not correspond to the peak area of the velocity-integrated intensity.

\begin{figure*}[t]
\begin{center}
\includegraphics[width=16cm, bb=0 0 1443 468]{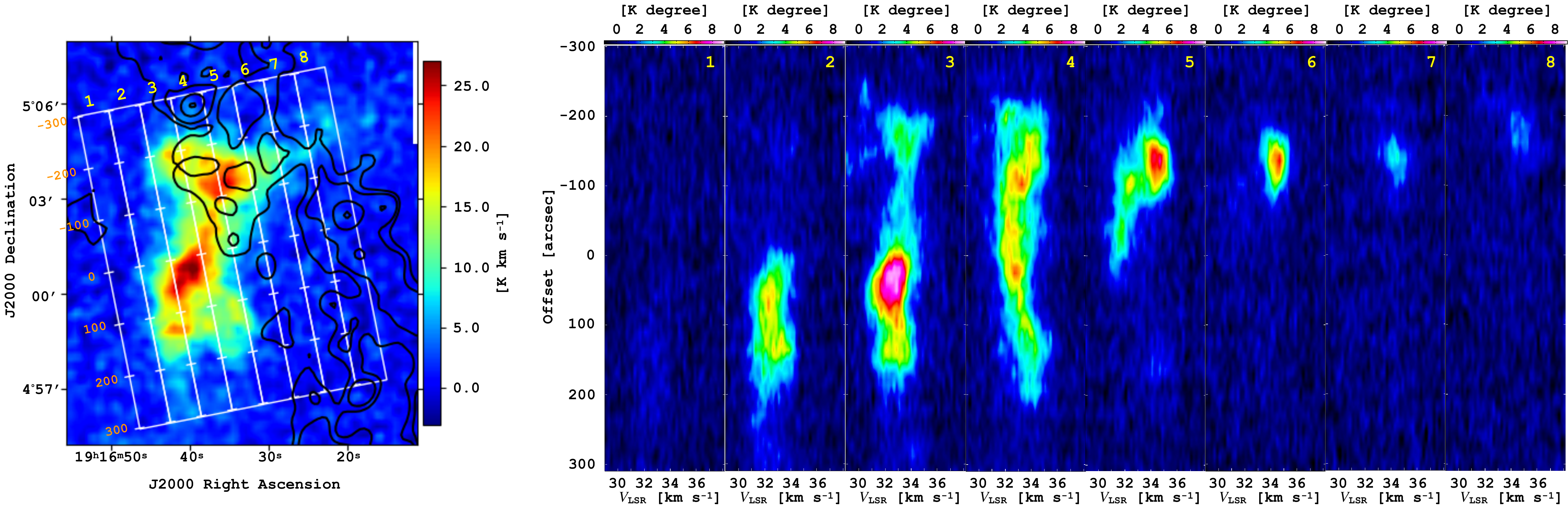}
\caption{Position--velocity diagrams of the $^{12}$CO($J$=1--0) emission of the chimney cloud. The intensity is integrated along the short axes of each square.}
\label{fig:pv_chimney}
\end{center}
\end{figure*}
\begin{figure*}[t]
\begin{center}
\includegraphics[width=15cm, bb=0 0 1318 456]{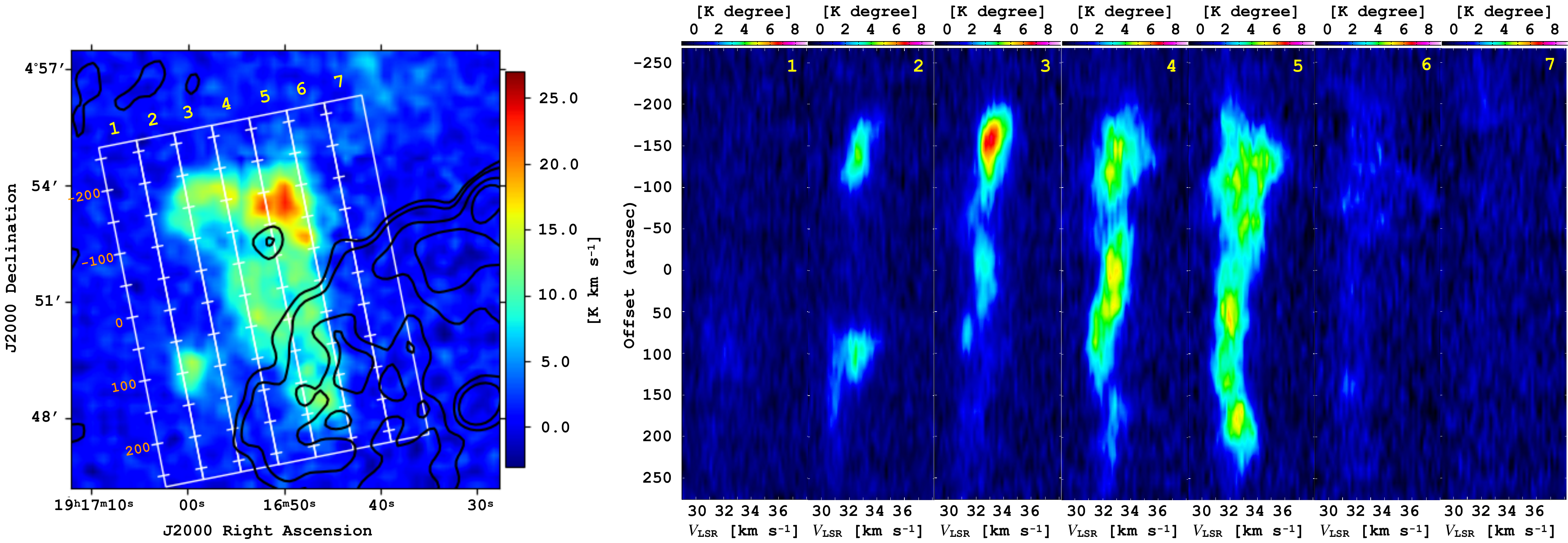}
\caption{Similar to figure \ref{fig:pv_chimney} but for the edge cloud.}
\label{fig:pv_edge}
\end{center}
\end{figure*}
\begin{figure*}[htbp]
\begin{center}
\includegraphics[width=16cm, bb=0 0 1160 760]{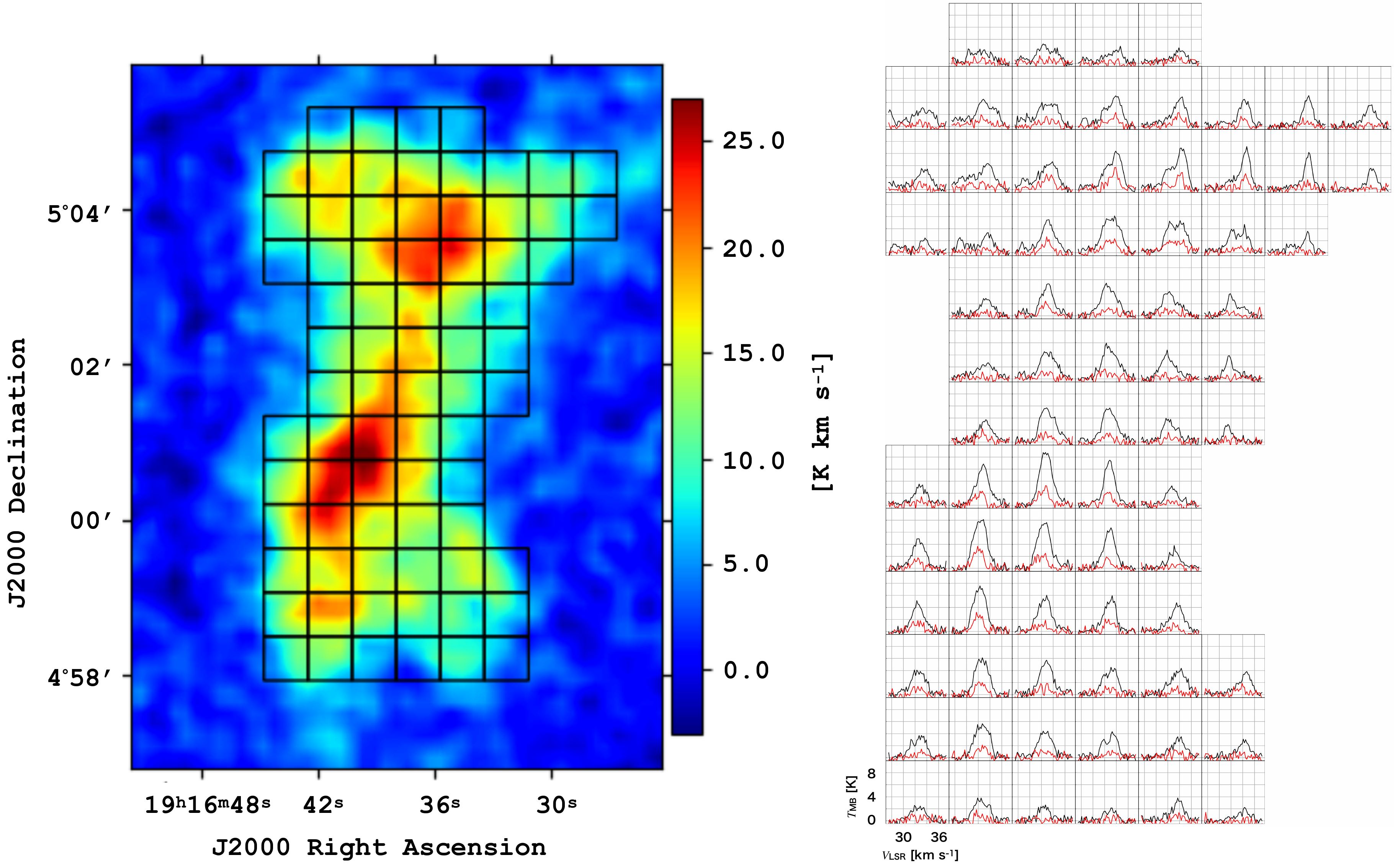}
\caption{(Left) Integrated intensity map of the $^{12}$CO($J$=1--0) emission of the chimney cloud. The velocity range is 30.1--36.5 km s$^{-1}$. (Right) $^{12}$CO($J$=1--0) (black) and three times $^{13}$CO($J$=1--0) (red) spectra for the regions shown in the grids in the left panel.}
\label{fig:profile_chimney}
\end{center}
\end{figure*}
\begin{figure*}[htbp]
\begin{center}
\includegraphics[width=16cm, bb=0 0 1067 660]{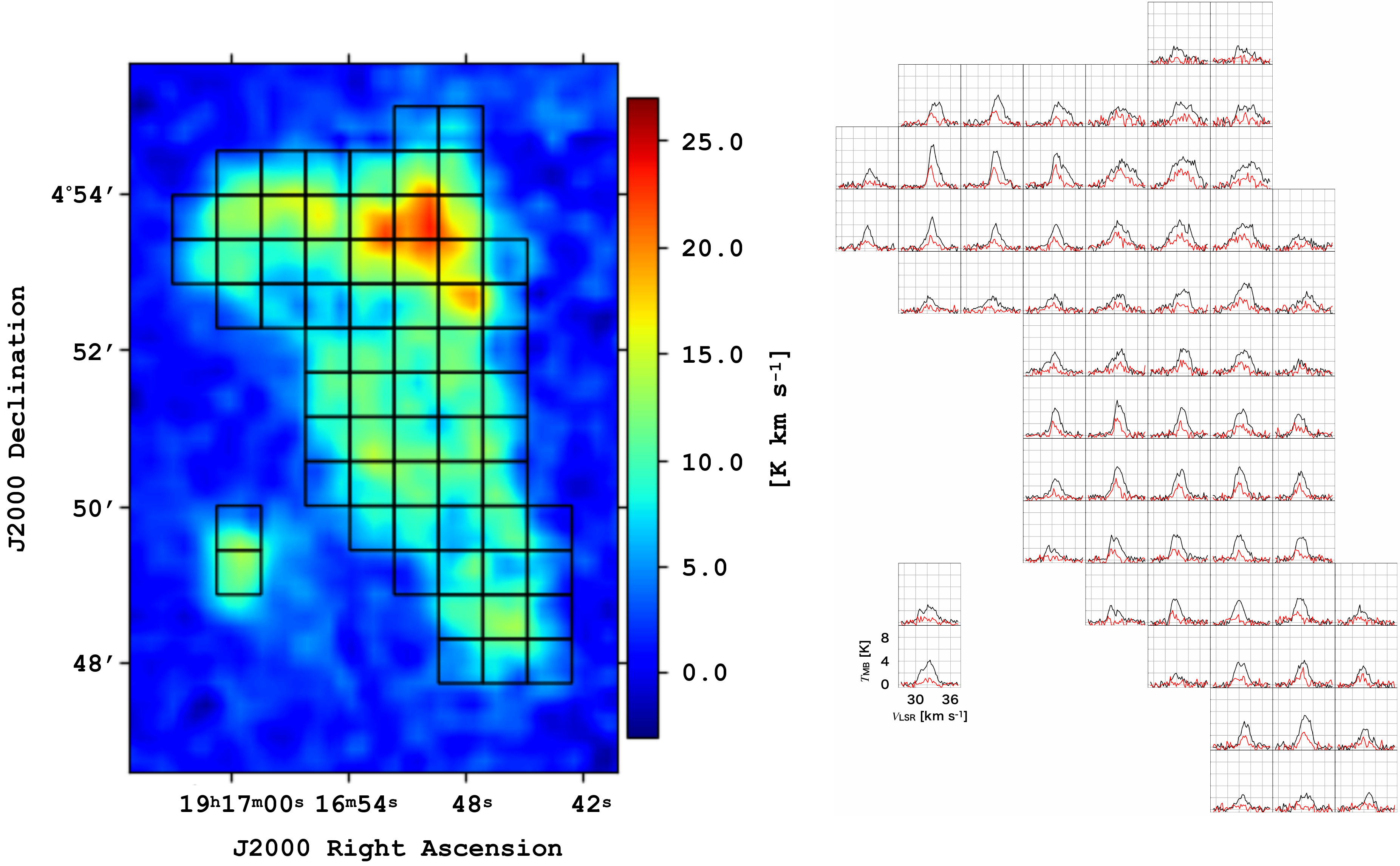}
\caption{Similar to figure \ref{fig:profile_chimney} but for the edge cloud.}
\label{fig:profile_edge}
\end{center}
\end{figure*}

\begin{figure*}[t]
\begin{center}
\includegraphics[width=16cm, bb=0 0 845 459]{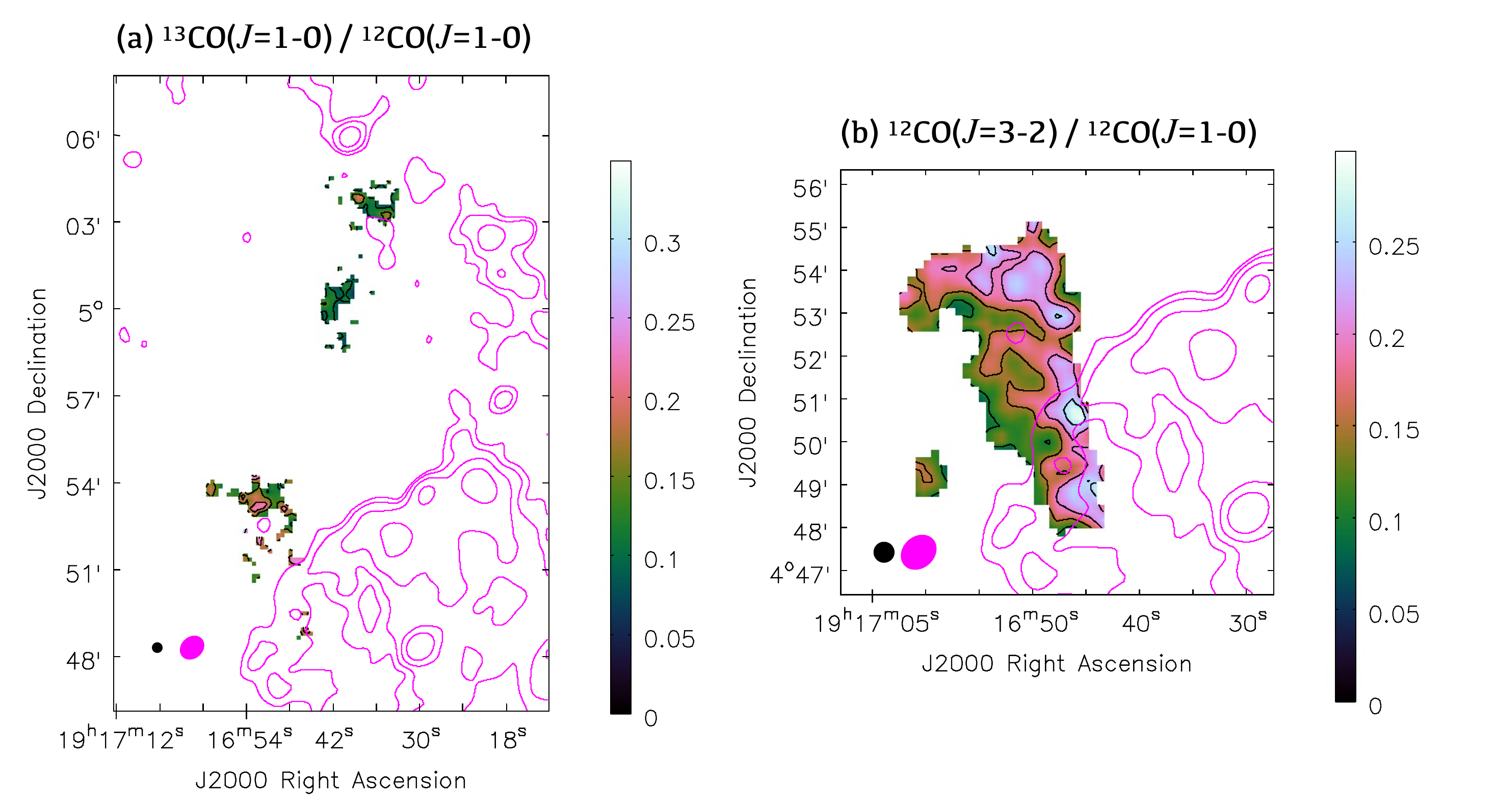}
\caption{Intensity ratios of $^{13}$CO($J$=1--0) and $^{12}$CO($J$=1--0) (a) and $^{12}$CO($J$=3--2) and $^{12}$CO($J$=1--0) (b). Only the regions where the intensities of the CO emission are higher than three times the rms value are plotted. Black contours show the values of [0.1, 0.15, 0.2, 0.25]. Magenta contours show the radio continuum observed with the JVLA at 1.602 GHz. The resolutions of intensity ratio maps and continuum are given at the bottom-left corner of each panel by black circle and magenta ellipse, respectively.}
\label{fig:intensity-ratios}
\end{center}
\end{figure*}
\begin{figure*}[t]
\includegraphics[width=16cm, bb=0 0 862 991]{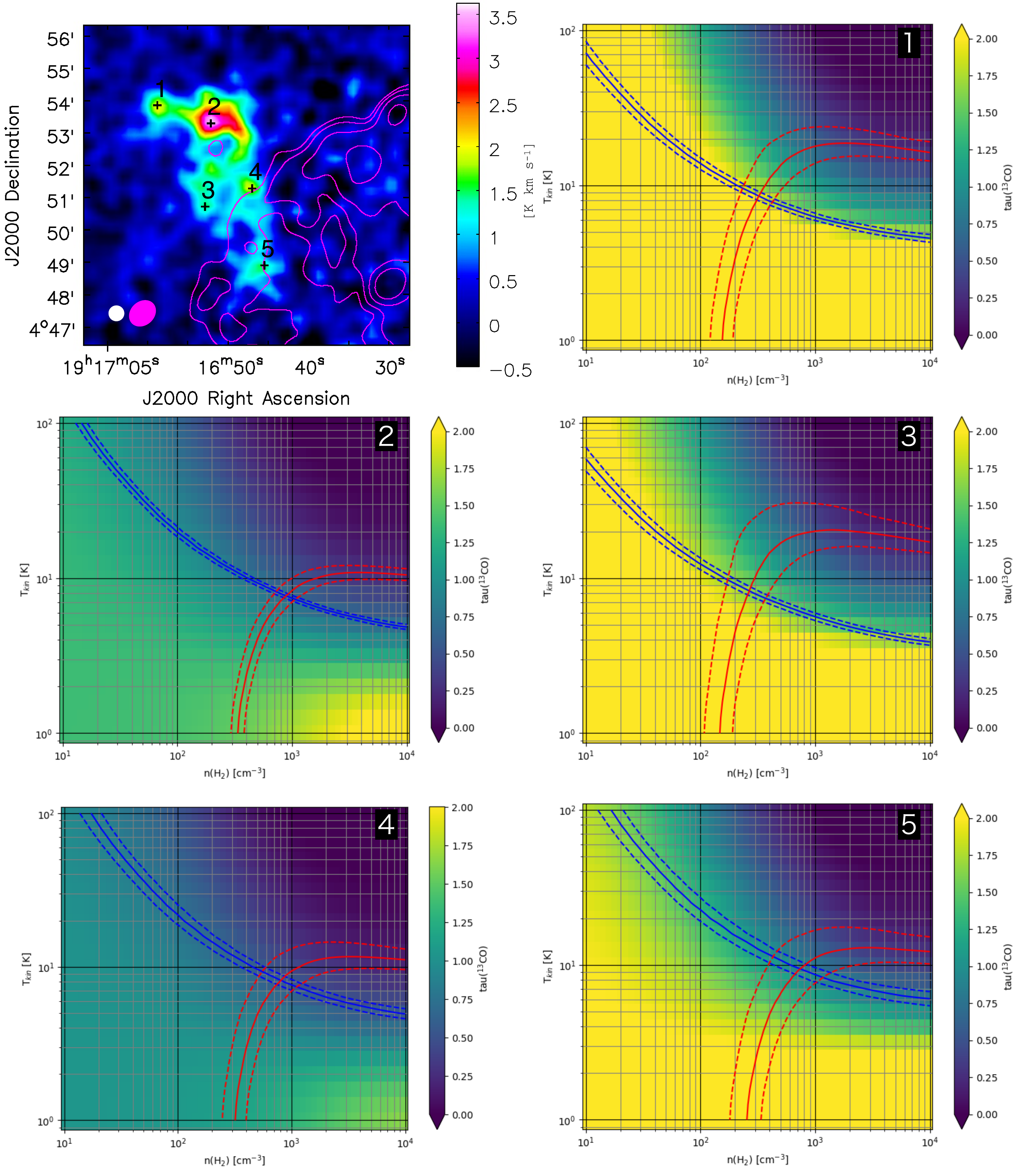}
\caption{Integrated intensity map of the $^{12}$CO($J$=1--0) emission of the chimney cloud in the velocity range of 30.1--36.5 km s$^{-1}$ (top-left) and the results of RADEX analysis conducted at the points shown on the map. Red and blue solid lines respectively represent the intensity ratios $^{13}$CO($J$=1--0)/$^{12}$CO($J$=1--0) and $^{12}$CO($J$=3--2)/$^{12}$CO($J$=1--0) at each position. The dashed lines associated with each solid line show the 1$\sigma$ of the ratio. The color represents the optical depth $\tau(^{13}{\rm CO})$. The parameters at the crossover points of the line intensity ratios correspond to the physical values derived by the RADEX calculation.}
\label{fig:LVG_edge}
\end{figure*}

\begin{figure*}[t]
	\includegraphics[width=16cm, bb=0 0 1215 519]{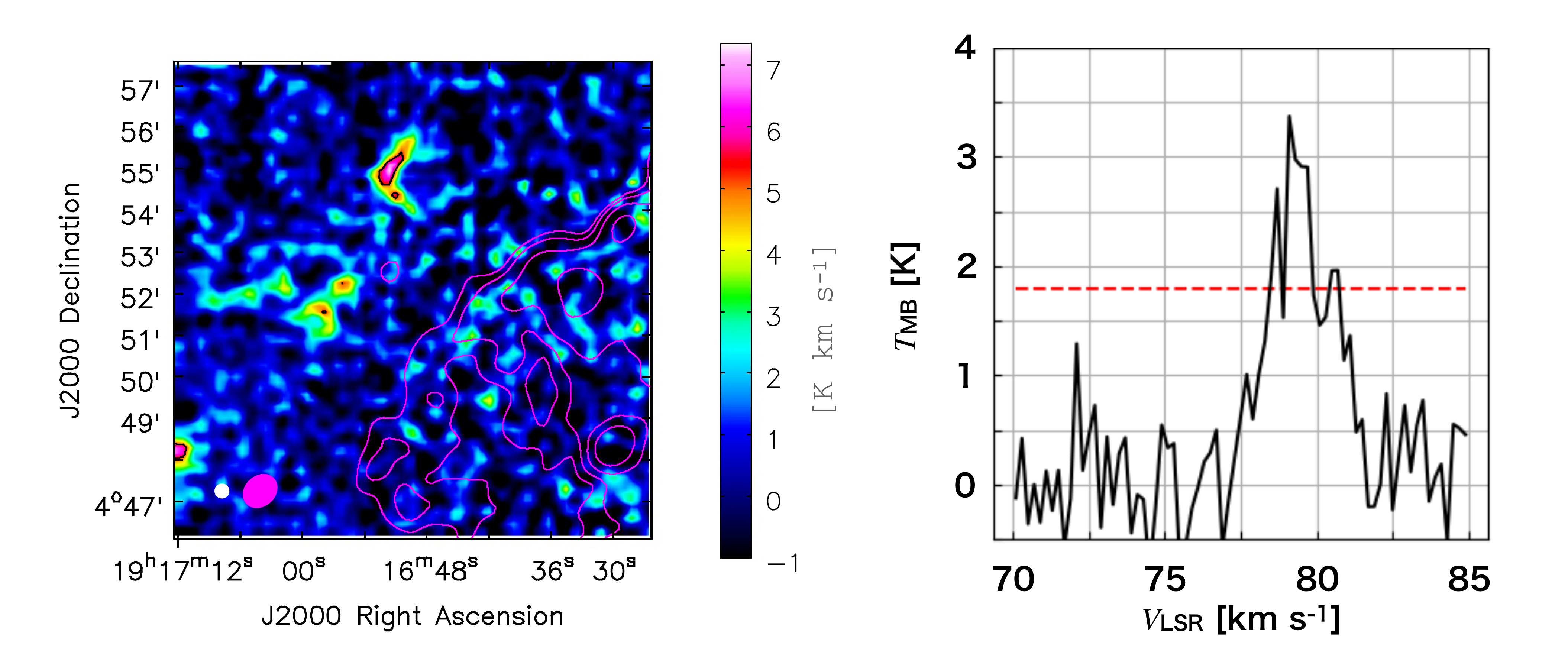}
    \caption{(Left) Velocity-integrated intensity maps of $^{12}$CO($J$=1--0). The velocity range is 77.7--83.7 km s$^{-1}$. Black contours show the line of five times the rms value. Magenta contours show the radio continuum observed with the JVLA at 1.602 GHz. The beam sizes of CO emission and continuum observations are shown in the bottom-left corner of each panel by a white circle and magenta ellipse, respectively. (Right) $^{12}$CO($J$=1--0) spectra of the peak position of the high-velocity clouds shown in left panel. The red-dashed line shows the value of three times the rms.}
    \label{fig:high-vel}
\end{figure*}

\subsection{Velocity structures}
We next explain the velocity structures of the molecular clouds in the eastern region of W50. Figure \ref{fig:pv_chimney} is a position--velocity diagram of the chimney cloud. Note that we inclined the boxes to align the integration axis with the direction of the SS 433 jet axis and thus assess the effect of the jet activity. We divided the chimney cloud into eight regions and investigated the velocity structures in the northeast-to-southwest direction. The most outstanding feature is seen in region 4, where the chimney cloud seems to touch the chimney of W50. In this region, there is a curved structure of the position--velocity diagram. The center velocity seems to shift approximately 1 km s$^{-1}$ along the curve. A similar trend is seen for the edge cloud (figure \ref{fig:pv_edge}). We divided the edge cloud into seven regions. The curved structures are in regions 4 and 5. Especially in region 5, the curved structure is above an offset of 0 arcsec, where the edge cloud also touches the eastern edge of W50 in the sky plane. Additionally, the velocity width is thick ($\sim$ 4 km s$^{-1}$) at an offset less than 0 arcsec in region 5.

Figure \ref{fig:profile_chimney} presents spectra for different parts of the chimney cloud. At the northern peak, the spectrum tends to have a wide velocity width and seems to have a spectral wing toward lower velocity. Between the northern and southern peaks, spectral wings form toward higher velocity. Spectral wings often imply the existence of an external force generated by surrounding sources. However, these wings may originate from multiple clouds in the line of sight and at slightly different velocities. Figure \ref{fig:profile_edge} presents spectral plots of the edge cloud. At the peak positions in the northwest region, we clearly see a wide spectrum not only for $^{12}$CO($J$=1-0) but also for $^{13}$CO($J$=1-0) with a mean velocity width of 4.26 km s$^{-1}$, which is a result similar to that in figure \ref{fig:pv_edge}. In the northeast region of the edge cloud, spectral wings form toward high velocity.

\subsection{Intensity ratios}
Panel (a) of figure \ref{fig:intensity-ratios} shows the intensity ratio of $^{13}$CO($J$=1--0)/$^{12}$CO($J$=1--0), which traces high-density areas of molecular clouds. Most parts of the chimney and edge clouds have values below 0.3. In the edge cloud, we see again the patchy structure, and we analyze the physical properties of each clump in section \ref{sec:discussion}. Panel (b) of figure \ref{fig:intensity-ratios} shows the intensity ratio of $^{12}$CO($J$=3--2)/$^{12}$CO($J$=1--0) of the edge cloud. There is a gradient from west to east, which suggests a difference in temperature and/or density in the western and eastern parts. 
\section{Discussion}
\label{sec:discussion}
 In this section, we consider the relation between the molecular clouds that we identified and their surroundings. We first derive the physical parameters of the chimney and edge clouds. We then consider the interaction between these clouds and the SS 433/W50 system. We also mention the relation between these clouds and a giant molecular filament in front of the SS 433/W50 system. We finally discuss the high-velocity clouds in terms of the relation with the clouds identified by \citet{su2018}.

\subsection{Physical properties of the molecular clouds}

We first estimated the column densities of each region shown in panel (b) of figure \ref{fig:integ_30.1-36.5} adopting two methods; i.e., assuming the X-factor and local thermodynamic equilibrium (LTE). The X-factor converts the $^{12}$CO($J$=1--0) integrated intensity $W(^{12}{\rm CO}(J=1$--$0))$ to the column density $N({\rm H}_{2})$, and we adopt the value of $N({\rm H}_{2})$/$W(^{12}{\rm CO}(J=1$--$0))$ = $2 \times 10^{20}$ cm$^{-2}$ (K km s$^{-1}$)$^{-1}$ with $\pm$ 30\% uncertainty \citep{bolatto2013}. Alternatively, we used the following procedures to derive the column density $N({\rm H}_{2})$ by assuming LTE \citep {wilson2009}. Assuming the $^{12}$CO($J$=1--0) line is optically thick, the excitation temperature $T_{\rm ex}$ is derived from the $^{12}$CO peak intensity $T_{\rm MB}$:
\begin{equation}
	T_{\rm ex} = 5.5\left[\ln{\left(1+\frac{5.5}{T_{\rm MB}+0.82}\right)}\right]^{-1}\ {\rm [K]}.
	\label{eq:excitation_temperature}
\end{equation}
The optical depth $\tau_{13}(v)$ is calculated for the $^{13}$CO($J$=1--0) brightness temperature of each velocity channel $T_{\rm MB}(v)$ as
\begin{equation}
	\tau_{13}(v)= -\ln{\left[1-\frac{T_{\rm MB}(v)}{5.3}\left\{\frac{1}{\exp{\left(\frac{5.3}{T_{\rm ex}}\right)}-1}-0.16\right\}^{-1}\right]}.
	\label{eq:tau13}
\end{equation}
Using the velocity resolution $\Delta v$ = 0.2 km s$^{-1}$, the column density of $^{13}$CO, $N(^{13}{\rm CO})$, is calculated as
\begin{equation}
	N(^{13}{\rm CO}) = 2.4 \times 10^{14} \times \sum_{v}\frac{T_{\rm ex}\tau_{13}(v)\Delta v}{1-\exp{\left(-\frac{5.3}{T_{\rm ex}}\right)}}\ [{\rm cm}^{-2}].
\end{equation}
We finally adopt the conversion factor from $N(^{13}{\rm CO})$ to $N({\rm H}_{2})$ of $7.7 \times 10^{5}$ \citep{frerking1982, pineda2010,wilson1994}.
We present the results in table \ref{tab:column_densities}. The column densities derived by assuming the X-factor are slightly higher than those derived using the LTE; however, we can explain these differences by errors in the observations and the abundance ratios [$^{13}$CO]/[H$_{2}$] = 1--3.5 $\times$ 10$^{-6}$ and the X-factor = 1.8--2.0 $\times$ 10$^{20}$ cm$^{-2}$ (K km s$^{-1}$)$^{-1}$ depending on the surrounding environment. 
We calculated the masses of the chimney and edge clouds using $N({\rm H}_{2})$ derived by assuming the X-factor,
\begin{equation}
	M_{\rm mol} = \bar{\mu} m_{\rm H} \sum{[N({\rm H}_{\rm 2})\times \Omega \times D^{2}]},
	\label{eq:mass}
\end{equation}
where $\bar{\mu}=2.8$, $m_{\rm H}=1.67\times10^{-24}$ g, $\Omega=1.7\times10^{-9}$ sr, and $D$ pc are the mean molecular weight, the mass of the atomic hydrogen, the solid angle subtended by the grid spacing, and the distance to the molecular clouds, respectively. If we assume the distance of $D=5500$ pc, the masses of the chimney and edge clouds are 3800 and 2300 $M_{\odot}$, respectively. Also, the masses of these clouds are 1100 and 700 $M_{\odot}$ assuming the distance of $D=3000$ pc.
 \begin{table}
 \tbl{Column densities\footnotemark[$*$]}{%
\begin{tabular}{lcc}
 \hline
  \multicolumn{1}{c}{Region} & $N^{1-0}_{\rm X}$ & $N^{13,1-0}_{\rm LTE}$\\
 \hline
  chimney 1 & 3.18 & 2.28 \\
  chimney 2 & 4.14 & 2.44\\
  chimney 3 & 5.27 & 2.70 \\
  chimney 4 & 4.72 &  2.14\\
  edge 1 & 2.66 & 1.90 \\
  edge 2 & 3.68 & 2.97 \\
  edge 3 & 2.82 & 1.40 \\
  edge 4 & 2.03 & 1.72 \\
  edge 5 & 2.08 & 1.64 \\ 
 \hline
\end{tabular}}\label{tab:column_densities}
\begin{tabnote}
 \footnotemark[$*$] Column 1: Region name given in panel (b) of figure \ref{fig:integ_30.1-36.5}. Column 2: Column density of H$_{\rm 2}$ derived from $^{12}$CO(J=1--0) in 10$^{21}$ cm$^{-2}$ assuming the X-factor. Column 3: Column density of H$_{\rm 2}$ derived from $^{13}$CO(J=1--0) in 10$^{21}$ cm$^{-2}$ assuming LTE.
\end{tabnote}
\end{table}

\begin{table*}
 \tbl{Results of RADEX calculation.\footnotemark[$*$]}{%
\begin{tabular}{lccccc}
 \hline
  \multicolumn{1}{c}{Region} & $R^{13/12}_{1-0}$ & $R^{12}_{3-2/1-0}$ & $\tau (^{13}{\rm CO})$ & $n({\rm H}_{\rm 2}$) & $T_{\rm kin}$\\
 \hline
  edge 1 & 0.154$\pm$0.028 & 0.190$\pm$0.020 & 1.87$_{-0.12}^{+0.17}$ & 320$_{-100}^{+120}$ & 8.6$_{-1.2}^{+1.6}$\\
  edge 2 & 0.195$\pm$0.020 & 0.201$\pm$0.014 & 0.72$_{-0.04}^{+0.04}$ & 880$_{-180}^{+220}$ & 7.7$_{-0.6}^{+0.8}$\\
  edge 3 & 0.123$\pm$0.028 & 0.137$\pm$0.018 & 1.64$_{-0.10}^{+0.13}$ & 270$_{-90}^{+110}$ & 8.3$_{-1.3}^{+2.2}$\\
  edge 4 & 0.204$\pm$0.040 & 0.228$\pm$0.030 & 0.56$_{-0.06}^{+0.07}$ & 800$_{-280}^{+400}$ & 8.2$_{-1.2}^{+1.8}$\\
  edge 5 & 0.166$\pm$0.040 & 0.244$\pm$0.033 & 0.93$_{-0.16}^{+0.18}$ & 740$_{-340}^{+660}$ & 9.5$_{-1.8}^{+2.8}$\\ 
 \hline
\end{tabular}}\label{tab:radex_results}
\begin{tabnote}
 \footnotemark[$*$] Column 1: Region name given in figures \ref{fig:integ_30.1-36.5} and \ref{fig:LVG_edge}. Column 2: Intensity ratio of $^{13}$CO($J$=1--0)/$^{12}$CO($J$=1--0). Column 3: Intensity ratio of $^{12}$CO($J$=3--2)/$^{12}$CO($J$=1--0). Column 4: Optical depth of $^{13}$CO. Column 5: Number density of H$_{\rm 2}$ in cm$^{-3}$. Column 6: Kinematic temperature in K.
\end{tabnote}
\end{table*}

Additionally, we carried out the RADEX calculation to estimate the temperature and density of the edge cloud using non-local thermodynamic equilibrium radiative transfer code \citep{vandertak2007}. We input parameter sets of $T_{\rm k}$ and $n(H_{\rm 2}$) for intervals of 10$^{0.1}$ K in the range of 1 to 100 K and intervals of 10$^{0.05}$ cm$^{-3}$ in the range of 10 to 10$^{4}$ cm$^{-3}$, and we calculated the intensities $^{12}$CO($J$=1--0), $^{13}$CO($J$=1--0) and $^{12}$CO($J$=3--2). The line intensity ratios of $^{13}$CO($J$=1--0)/$^{12}$CO($J$=1--0) and $^{12}$CO($J$=3--2)/$^{12}$CO($J$=1--0) were then calculated in $n(H_{\rm 2}$)--$T_{\rm k}$ space. We estimated $N(^{12}$CO) as
\begin{equation}
	N(^{12}{\rm CO})=N({\rm H_2}) \times 10^{-4},
	\label{eq:12co-density}
\end{equation}
where $N({\rm H}_{2})$ was derived by assuming the X-factor. Additionally, we assumed $N(^{12}$CO)/ $N(^{13}$CO)=62 \citep{milam2005}.
We analyzed the local peaks of $^{13}$CO($J$=1--0) of the edge cloud shown in the top-left panel of figure \ref{fig:LVG_edge}. The observed line intensity ratios of $^{13}$CO($J$=1--0)/$^{12}$CO($J$=1--0) and $^{12}$CO($J$=3--2)/$^{12}$CO($J$=1--0), with one sigma error, are plotted in $n(H_{\rm 2}$)--$T_{\rm k}$ space with red and blue lines in five plots of figure \ref{fig:LVG_edge} with the expected optical depth $\tau (^{13}{\rm CO})$. The parameters at the crossover points of the line intensity ratios correspond to the physical values derived by the RADEX calculation and are listed in table \ref{tab:radex_results}. We note that the densities of all regions are much lower than the critical density of $^{13}$CO($J$=1--0); nevertheless, the $^{13}$CO($J$=1--0) emission is observed. This implies that the edge cloud comprises tiny clumps that cannot be spatially resolved by our observations and are distributed with the low beam-filling factor. In that case, the emission is smoothed out, and the densities are estimated to be lower. Additionally, we find that the western side of the edge cloud has higher density. This result is consistent with the trend of the intensity ratio $^{12}$CO($J$=3--2)/$^{12}$CO($J$=1--0) shown in panel (b) of figure \ref{fig:intensity-ratios}.


\subsection{Association with the SS 433/W50 system}
We here consider the relation between the chimney and edge clouds and the SS 433/W50 system. Our observations revealed that these clouds are close to the eastern region of W50 in the plane of the sky. The clouds have clumpy spatial structures and complex velocity structures, including a shift in the central velocity (figures \ref{fig:pv_chimney} and \ref{fig:pv_edge}), spectral wings, and wider velocity widths (figures \ref{fig:profile_chimney} and \ref{fig:profile_edge}). Additionally, we identified that the density distribution of the edge cloud is not flat (panel b of figure \ref{fig:intensity-ratios} and figure \ref{fig:LVG_edge}). Near the chimney and edge clouds, there is no heating source that might explain their complexity, such as the HII region radiating UV photons. Also, the gamma-ray emission regions of the SS 433 jet are far from these clouds, above 0.6$^{\circ}$, and should have less influence. Therefore, the properties of these clouds imply the interaction with the eastern ear of W50.

Unfortunately, we cannot conclude the interaction from only our observations. There is a possibility that the clouds are mere foreground or background sources of W50. As an example, \citet{lin2020} identified a giant molecular filament in the Milky Way Imaging Scroll Painting (MWISP) in front of the SS 433/W50 system in the velocity range from 27 to 40 km s$^{-1}$, named GMF MWISP G041-01. We thus need to discuss the association between the molecular clouds that we identified and this giant molecular filament. \citet{lin2020} suggested that GMF MWISP G041-01 comprises four components, three being filamentary structures. Two such giant filaments possibly collided in the region around ($\alpha_{J2000}$, $\delta_{J2000}$) = (19$^{h}$ 10$^{m}$ 01.3525$^{s}$, +7$^{d}$ 15$^{m}$ 03.330$^{s}$) and (19$^{h}$ 10$^{m}$ 29.6389$^{s}$, +6$^{d}$ 10$^{m}$ 32.713$^{s}$), while they might be just overlapping along the line of sight. If such a collision actually occurred, spectral wings, such as those observed for the chimney and edge clouds, should be seen. The clouds that we identified might then be part of GMF MWISP G041-01. However, the filament--filament collision is located away from the clouds, by approximately 2.4 degrees, corresponding to 71 pc assuming the distance of GMF MWISP G041-01 to be 1.7 kpc. We thus infer that the clouds that we identified are independent of the system of GMF MWISP G041-01.

Hereafter, we assume that the chimney and edge clouds are related to W50 and consider the formation mechanism of these molecular clouds. An instinctive formation scenario is that the eastern ear of W50 compressed the surrounding HI gas, and a shock induced the molecular cloud formation \citep{asahina2014,asahina2017}. However, this mechanism would take a long time, approximately $10^{6}$ yr, to form molecular clouds, and this time is in disagreement with the age of W50 (a few $10^{4}$ yr). We then consider the alternative scenario of sweeping tiny molecular clumps by the surface of the eastern ear of W50, corresponding to the surface of the jet cocoon. As mentioned above, the chimney and edge clouds seem to comprise more small clumps. Additionally, we identified faint and clumpy clouds in the eastern region of W50 in the velocity range of 30.1--42.3 km s$^{-1}$ except for the chimney and edge clouds (figure \ref{fig: integ_wide-vel}). This means that many tiny clumps are drifting at the position of the eastern ear of W50. The origin of the tiny clumps remains unknown, but there is a possibility that they were formed by the activity of the progenitor star of SS 433, such as by the compression of the surrounding HI gas by the stellar wind. Here, we imagine that the faint clouds are distributed throughout the region before the eastern ear goes through. The mean column density of H$_{2}$ in the velocity range of 30.1--42.3 km s$^{-1}$ except for the chimney and edge clouds is 7.18 $\times$ 10$^{20}$ cm$^{-2}$. We regard the eastern ear as a cylinder with a diameter of 0.3 deg and a height of 0.7 deg. Assuming a distance of 3.0 and 5.5 kpc, the eastern ear can collect clumps to a mass of 7.2 $\times$ 10$^{3}$ M$_{\odot}$ and 2.4 $\times$ 10$^{4}$ M$_{\odot}$, respectively. If we reduce the velocity range and consider only the contribution of the northern part of the eastern ear, the mass is comparable to that of the total mass of the chimney and edge clouds. Additionally, this scenario explains the complexity of the velocity structures of the clouds. Note that these clumps might influence the formation theory of the eastern ear of W50 \citep{goodall2011,ohmura2021}. To consider how they affect, we require to carry out additional numerical simulations; however, it is out of the focus of this paper. 

Also, we discuss the distance of the SS 433/W50 system based on the relation with the newly identified clouds. Using the velocity of 33 km s$^{-1}$, roughly the peak velocity of the chimney and edge clouds, the kinematic distance is 2.2 kpc. This value is far from the historically discussed one in the range of 3.0--5.5 kpc. Here, we suggest that the clouds are unsuitable for deriving the distance based on the model for Galactic rotation. As mentioned above, these clouds probably consist of many small clumps, and they are affected by the motion of the SS 433 jet. Small clumps can be easily shifted the central velocity. Therefore, there is a possibility that these clouds do not follow simply the rotation of the galactic plane. This scenario can explain the reason why the velocities of these clouds are lower shifted; since the eastern jet is approaching, the clouds have been pushed toward us.

We mention the differences between the clouds we identified and reported by \citet{yamamoto2022}, N4 (see figure \ref{fig:W50} and table \ref{table:prev-MC}). N4 is not similar to the typical molecular clouds in the Galactic plane, and it shows clear intensity and velocity gradients and a velocity shift from the systemic. In addition, the kinematic temperature of a part of N4 is as high as $\sim$ 50 K. The authors suggested that some external force is required to explain these trends. Considering the sources around N4, the SS 433 jet is the most plausible and is thought to be colliding with the cloud. In fact, the X-ray jet and N4 overlap in the plane of the sky, and this scenario is reasonable. On the other hand, although the edge cloud we identified seems to have a density gradient, the absolute values are low compared to the typical Galactic clouds. Also, the cloud has the typical kinematic temperature. The difference between N4 and the newly identified clouds may come from the difference in impact given by the jet. N4 is closer to SS 433 and its jet axis, and the active jet is interacting with N4. On the other hand, the clouds we identified are far from SS 433 and the jet axis, and they are interacting with the surface of the eastern ear of W50, corresponding to the surface of the jet cocoon. Additional observations such as shock tracer should verify the difference of influence of the SS 433 jet on these molecular clouds.


\subsection{High-velocity clouds in the eastern region of W50}
Finally, we mention molecular clouds in a higher velocity range. The left panel of figure \ref{fig:high-vel} is a velocity-integrated intensity map of $^{12}$CO($J$=1--0) at a velocity of 77.7--83.7 km s$^{-1}$. As mentioned in section \ref{sec:intro}, \citet{su2018} suggested that the clouds in this velocity range relate to the SS 433/W50 system, and we also identified faint clumps with our observation in the closer region of W50. Since they have completely different velocities, these high-velocity clouds should be independent structures from the chimney and edge clouds. We show the spectrum of the peak intensity position in right panel of figure \ref{fig:high-vel}. The center velocity and velocity width are 79.4 km s$^{-1}$ and 2.41 km s$^{-1}$, respectively.

In fact, the clumps that we identified seem to be similar to the molecular clouds reported by \citet{su2018}. However, they have a different feature (see figure 13 of their paper). \citet{su2018} suggested that the eastern molecular clouds of the SS 433/W50 system are approaching us. In their model, the clouds closer to the SS 433/W50 system have a higher velocity owing to their interaction with the jet. Although the clumps that we identified are closer to W50 than their clouds, the central velocity is lower than their highest cloud velocity of 84 km s$^{-1}$. This means that the high-velocity clouds in the present study are not a member of the clouds identified by \citet{su2018}, or they have slightly different kinematic properties owing to the offset from the SS 433 jet axis. Note that the detected signal from these clumps is very weak, and our observation is not suitable for discussing the details. We require to observe them deeper to extract more robust conclusion.

\section{Conclusion}
\label{sec:conclusion}
We reported the observation of molecular clouds at the eastern edge of W50 with the Nobeyama 45-m telescope and the ASTE. We identified two clouds that possible interact with the SS 433/W50 system for the first time. One is in the northern region of the eastern ear of W50, where there is a protruding structure called the chimney. The other is located at the eastern edge of the ear. Both clouds comprise small clumps that might not be resolved sufficiently by our observations. They have complex velocity structures with spectral wings and broadening, and it is unclear whether these features are due to an interaction with W50 or merely overlapping multiple components. The distribution of the intensity ratio and the results of RADEX analysis reveal that the western side of the edge cloud has higher density. Although it is difficult to conclude the relationship between the clouds that we identified and the SS 433/W50 system, the clouds were possibly formed by the propagation of the eastern ear of W50 and the sweeping of tiny clumps. Finally, we identified the high-velocity clouds at the eastern edge of W50 in the same velocity range as the clouds reported by \citet{su2018}, while it is unclear whether they belong to the same series.
\section*{Acknowledgements}
We are grateful to Drs. M. Kohno, K. Tsuge, Y. Yamane and Mr. D. Tsutsumi for supporting our observations. We thank Dr. H. Sano for helpful comments and discussion for the interaction of the cloud with the GXB jet. We thank the anonymous referee for useful comments and constructive suggestions. This work was supported by JSPS KAKENHI Grant Numbers HS: 20J13339, 22K20386, MM: 19K03916, 20H01941, 22H01272, and KT: 18H05440,
20H01945, 22H00152. The Nobeyama 45-m radio telescope is operated by Nobeyama Radio Observatory, a branch of National Astronomical Observatory of Japan. The ASTE telescope is operated by National Astronomical Observatory of Japan (NAOJ). The National Radio Astronomy Observatory is a facility of the National Science Foundation (NSF) operated under a cooperative agreement by Associated Universities, Inc. Data analysis was partly carried out on the Multi-wavelength Data Analysis System operated by the Astronomy Data Center (ADC) at the National Astronomical Observatory of Japan. This work was supported by the Japan Foundation for Promotion of Astronomy. SAOImageDS9 development was made possible by funding from the Chandra X-ray Science Center (CXC), the High Energy Astrophysics Science Archive Center (HEASARC), and the JWST Mission office at the Space Telescope Science Institute \citep{joye2003}. This research used Astropy,\footnote{http://www.astropy.org} a community-developed core Python package for Astronomy \citep{astropy2013, astropy2018}. We thank Edanz (https://jp.edanz.com/ac) for editing a draft of this manuscript.

\end{document}